# Clustering with Prototype Extraction for Census Data Analysis

O. Chertov

*Applied Mathematics Department*
*NTUU "Kyiv Polytechnic Institute"*
*Kyiv, Ukraine*

chertov@i.ua

M. Aleksandrova

*Applied Mathematics Department*
*NTUU "Kyiv Polytechnic Institute"*
*Kyiv, Ukraine*

rita.v.aleksandrova@gmail.com

*Abstract* – **Not long ago primary census data became available to publicity. It opened qualitatively new perspectives not only for researchers in demography and sociology, but also for those people, who somehow face processes occurring in society.**

**In this paper authors propose using Data Mining methods for searching hidden patterns in census data. A novel clustering-based technique is described as well. It allows determining factors which influence people behavior, in particular decision-making process (as an example, a decision whether to have a baby or not). Proposed technique is based on clustering a set of respondents, for whom a certain event have already happened (for instance, a baby was born), and discovering clusters' prototypes from a set of respondents, for whom this event hasn't occurred yet. By means of analyzing clusters' and their prototypes' characteristics it is possible to identify which factors influence the decision-making process. Authors also provide an experimental example of the described approach usage.**

## I. Introduction

Nowadays there exist two considerable global tendencies. The first trend is that the amount of digital information yearly produced by humanity increases significantly. Thus, in 2009 it grew by 62% and comprised almost 800.000 petabytes; in 2010 it is believed to reach almost 1.2 million petabytes [1]. The second trend is related to the fact that not only aggregated (consolidated) data but also primary ones become more available. Researchers can relatively easy get access to the great variety of primary data such as information on patients' hospital treatment (so-called Clinical Data Repositories [2]), electronic commerce results in big automated collections of consumer data, microfiles with large census data samples etc.

Usually, the access is given not to the complete primary data sets but to the data samples (microfiles). In addition, some attribute's values are masked somehow, or even absolutely unavailable. These restrictions are necessary to provide the published data anonymity, which is a subject to special legal regulation in different countries (e.g., see the Health Insurance Portability and Accountability Act of 1996 (HIPAA) [3]; the Patient Safety and Quality Improvement Act of 2005 (PSQIA) [4] concerning the health data protection in the USA; Directive on privacy and electronic communications [5] about electronic commerce in the EU; the State Statistics Law [6] about providing confidentiality of the primary statistical information in Ukraine).

However, the amount of available primary data keeps growing. The IPUMS project is the most significant example. It started in 1992 [7] with the main goal to collect and distribute census data for different researchers from around the world. Within it, more than 326 million person records collected from 159 censuses held in 55 countries (at the moment this paper is being written) are accessible for the researchers. The social importance of results, which can be obtained through such data analysis, can scarcely be overestimated. Still, such analysis requires using appropriate tools.

## II. Related Works

### A. Analysis Methods for Census Data

Statistical and OLAP databases are the main sources of population census data [8]. Different statistical methods are actively used for analyzing it, among them analysis of variance (ANOVA), regression analysis, log-linear analysis, nonparametric approaches [9]. While using statistical methods it is possible to achieve important results, but this approach requires prior knowledge or at least assumptions about some patterns existence.

As opposite to statistical methods intelligent data analysis (*Data Mining*) makes it possible to discover such patterns, which even wasn't suspected to exist. That is why Data Mining refers to extracting ("mining") knowledge from large amounts of data [10].

### B. Data Mining Techniques

Among the most widely used Data Mining techniques it is possible to define [11]: clustering (classifying), nearest neighbor prediction, decision trees, neural networks, and association rules.

Clustering is a process of grouping related points in a given set on the basis of having similar characteristics. Thus, clustering discovers natural accumulations in data sets. Some authors [12] also single out classification as clustering, where the analyst knows ahead of time how classes are defined.

Modern clustering algorithms can be divided in two groups: on-line and off-line methods [13]. On-line methods use every new data point for cluster centers identification.

Thus, the system is learning as adding new element. The other approach computes cluster centers only once and they can't be changed lately. Considering static nature of the census data it is logical to use off-line methods.

Commonly used off-line algorithms are *k*-means clustering, fuzzy *c*-means clustering, mountain clustering, and subtractive clustering [13]. First two algorithms require prior knowledge of clusters number. Thereby, they implement classification, not clustering. Both mountain and subtractive clustering techniques implement the same algorithm. The only major difference is that the first one examines every possible point in the data space (bounded by minimum and maximum values in each dimension) in order to discover cluster centers, and the second one goes only over points of the clustered data set. The latter approach significantly speeds up the algorithm's performance.

*C. Data Mining in Census Analysis*

Intelligent data analysis techniques are widely used in different scopes of human activities: education [14], medicine [15], banking [16], marketing [17], and so on. Still, Data Mining methods are almost not used for census data analysis. There are only few related works.

References [18-20] describe SPADA (*spatial pattern discovery algorithm*) system, which was designed for discovering spatial association rules in census data. Proposed method of rules discovery is based on a multi-relational data mining approach and uses representation and reasoning techniques developed in inductive logic programming.

CHARM, an algorithm for mining all frequent closed itemsets is proposed in [21]. CHARM is an alternative to Apriori-inspired algorithms [22, 23]. It was tested on census databases.

In [24] special requirements that occur for the aggregated spatial census data mining (*subgroups mining*) are discussed. Authors also describe subgroup mining system (SubgroupMiner) providing multi-relational hypotheses support, efficient data base integration, discovery of causal subgroup structures, and visualization based interaction options.

Thus all previous works concerning usage of Data Mining techniques in census analysis consider only spatial databases, which contain a part of census data. Unlike ideas proposed in mentioned papers, authors believe that methods of intelligent data analysis can be successfully used for discovering hidden patterns in primary census data (microfiles).

III. PROBLEM FORMULATION AND SOLUTION

*A. Microfile Structure*

A census microfile is a set of records of two types [25]: household records and person records. Each record contains certain attributes. For example, among household attributes there are state code, building size, number of rooms, type of unit (housing unit, institutional group quarters, noninstitutional group quarters) so on; person records contain such information as sex, age, marital status, race. All attributes are numerically coded.

Each person record has a field which determines this person's relationship with a householder. This field can take different values: "husband/wife", "natural born son/daughter", "uncle/aunt" so on.

*B. Paper Purpose*

The purpose of this paper is to develop an approach for identifying factors, which stimulate or vice versa do not contribute to the decision making process concerning respondents' life arrangement (for example, whether to study, have a baby, move to a new place).

During the experiment we analyzed population census microfile, as it contains all necessary socio-economical data. Clustering was chosen as an analysis technique, because it allows discovering natural data accumulations, thus, it gives an opportunity to identify which parameters consolidate certain people groups. We used subtractive clustering algorithm, which doesn't require prior knowledge of clusters number (note that in our case it is almost impossible to define clusters number beforehand) and is relatively fast [26]. Speed is particularly important thing while analyzing such large amounts of data as census data.

*C. Subtractive Clustering Algorithm*

Before using subtractive algorithm all data points must be rescaled in order to fall within a unit hypercube [27]. After this potential $P_i^{(1)}$ of each data point $x^i$ is calculated by (1).

$$P_i^{(1)} = \sum_{j=1}^{n} \exp\left(-\frac{\left\|x^i - x^j\right\|^2}{(r_a/2)^2}\right), \quad (1)$$

where $r_a$ is a positive constant called cluster radius, $n$ is a number of data-points in the clustered set.

A point with the highest potential is considered to be a first cluster center $c_1$. Then all potentials are recalculated by (2):

$$P_i^{(2)} = P_i^{(1)} - P_{c_1} \exp\left(-\frac{\left\|x^i - c_1\right\|^2}{(r_a \eta / 2)^2}\right), \quad (2)$$

where $P_{c_1}$ is a potential of the first cluster center, $\eta$ is a quash factor. By means of this recalculating influence of the first cluster center is excluded, as potentials of all close to it data points are reduced significantly.

Data point with the highest potential $P_k^{(2)}$ is tested to be the second cluster center.

In general, after the *m*-th cluster center has been obtained potentials of all data points are recalculated by (3):

$$P_i^{(m+1)} = P_i^{(m)} - P_{c_m} \exp\left(-\frac{\left\|x^i - c_m\right\|^2}{(r_a \eta / 2)^2}\right), \quad (3)$$

where $c_m$ is the $m$-th cluster center.

Point $c^*_{m+1}$ with the highest potential $P_l^{(m+1)}$ is accepted as an $(m+1)$-th cluster center if

$$P_l^{(m+1)} > \overline{\varepsilon} P_{c_1}, \quad (4)$$

where $\overline{\varepsilon}$ is an accept ratio – the potential of the cluster center candidate as a fraction of the first cluster center potential, above which data point $c^*_{m+1}$ is accepted as a cluster center.

Else if condition (5) holds point $c^*_{m+1}$ is rejected as a cluster center and the clustering process ends.

$$P_l^{(m+1)} < \underline{\varepsilon} P_{c_1}, \quad (5)$$

where $\underline{\varepsilon}$ is a reject ratio.

Otherwise the following condition is verified

$$\frac{d_{\min}}{r_a} + \frac{P_l^{(m+1)}}{P_{c_1}} \geq 1, \quad (6)$$

where $d_{\min}$ is the minimal distance between $c^*_{m+1}$ and all previously defined cluster centers.

If (6) holds $c^*_{m+1}$ is accepted as a cluster center and the algorithm starts next iteration. Else $c^*_{m+1}$ is rejected and its potential is set to 0; data point with the next highest potential on the current iteration is re-tested.

Quash factor, accept and reject ratios are specified by the user. Their default values in MATLAB framework are 1.25, 0.5 and 0.15 respectively. Cluster radius is also an input argument.

After all centers are defined the origin set can be divided into appropriate number of clusters by referring each point to that cluster, which center is the closest to the point under investigation.

*D. Novel Influence Search Algorithm Based on Clustering and Prototype Definition*

We propose a novel Influence search algorithm for solving a problem posed in the current section. This algorithm consists of the following steps.

1. Separate two groups $N_1$ and $N_2$ out of the origin dataset. The first group should contain records about those respondents who possess a certain characteristic, and the second one – records about respondents who do not possess it. Additional restrictions can be also imposed on the group definition process. They are usually a result of the problem domain analysis.
2. Identify those attributes, which can potentially influence the chosen characteristic presence. Mark out attributes for clustering, i.e., attributes which are numerical or can be compared by numbers.
3. Cluster group $N_1$.
4. Basing on the problem specific define invariant parameters for groups $N_1$ and $N_2$. Identify range of values for each parameter corresponding to the clusters' bounds.
5. Using obtained values ranges define clusters' prototypes out of the group $N_2$.
6. Compare characteristics of clusters and their prototypes; summarize results.

IV. EXPERIMENTAL RESULTS

The task of the experiment was to identify which factors influence human desire to have a baby.

Authors took a microfile with 5-percent sample of the California census data for 2000, which contains records about 610369 family households (we ignored subfamilies as number of households with subfamilies comprise only 3.6% of the initial sample) [25].

Following attributes were considered:
- home ownership,
- type of building,
- number of vehicles available,
- commercial business on property,
- parents' age,
- parents' education,
- parent's ancestry,
- class of worker for each parent,
- father's total income (in 1999).

Such parameter as mother's total income wasn't considered as a lot of women go on a maternity leave, thus, in our case this parameter can't be used as a family state factor.

We separated two groups out of the origin set of families. Group $N_1$ contains families with one or two children aged from 0 to 2 years, group $N_2$ is composed of families without children. Such restrictions on the children's age were imposed in order to track the change in family's state from childless to a family with a little child (children).

To obtain reliable results authors also imposed some additional restrictions on groups $N_1$ and $N_2$:
- all the families must be complete (presence of both father and mother is required);
- both parents must be without disabilities;
- parents' age must be within the most favorable period for having babies.

These conditions are quite relevant and obvious. For instance, it is clear that parents' illness affects significantly willingness and ability to have children.

In order to identify bounds of the most favorable age for having babies authors had to conduct additional researches. We calculated the number of families having a child or two children from 0 to 2 years depending on father's and mother's age. Obtained results are shown on Fig. 1.

From the diagrams it is clearly seen that the age of the first babies' bearing in families is almost normally distributed with average values 32 years for men and 30 years for women. As the interval of the most favorable age for having babies we took ranges 24-38 for men and 22-37 for women (values which correspond to 400 and more families). The size of the obtained sample is 3/4 of the initial set of families with a child or two children from 0 to 2 years.

Considering all imposed above restrictions we got groups $N_1$ and $N_2$ with 8299 and 12249 elements respectively.

The next step of the analysis was to identify parameters for clustering group $N_1$.

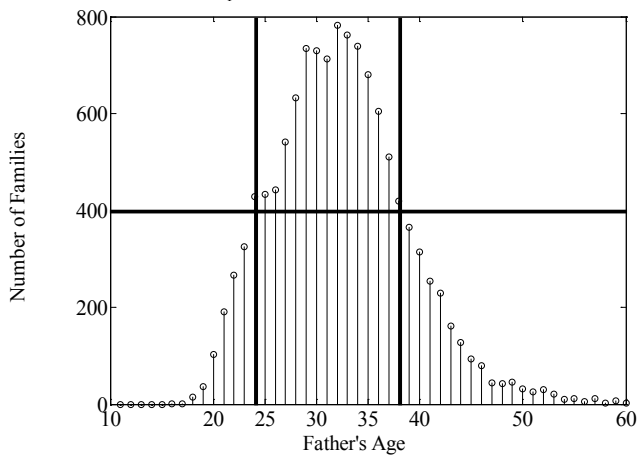

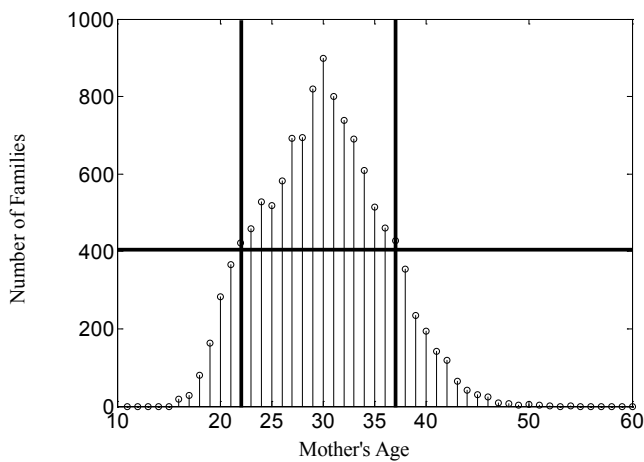

Fig. 1 Father's (a) and mother's (b) age distribution in families with little children.

We decided to use following parameters: father's age, mother's age, father's education, mother's education, total father's income. Such attributes as home ownership, class of worker and ancestry can't be used for clustering as they are qualitative, not quantitative characteristics and can be hardly compared by numbers.

Using subtractive algorithm for clustering group $N_1$ on the mentioned above attributes (we used default values of input arguments and 0.5 as a cluster radius) resulted in 3 clusters with centers given in Table I (education codes are presented in Table II).

Analyzing obtained cluster centers we can say that:
- cluster №2 contains the youngest couples with the lowest education level and the lowest incomes; relatively low income and education levels are partly a consequence of the fact that some parents are still studying;
- cluster №3 contains the oldest couples with the highest education and income levels;
- cluster №1 is comprised of those couples whose age is the closest to the expected values of the age distribution; education and income have intermediate values between 2-d and 3-d clusters;
- in each cluster parents have almost the same education levels.

TABLE I
CLUSTER CENTERS

| Clusters | 1 | 2 | 3 |
|---|---|---|---|
| Father's age | 32 | 27 | 36 |
| Mother's age | 30 | 26 | 34 |
| Father's education | 12 | 11 | 13 |
| Mother's education | 12 | 11 | 13 |
| Father's total income (in 1999), $ *10^4 | 5.0 | 3.5 | 6.7 |

TABLE II
EDUCATION CODES

| 1 | No schooling completed | 9 | High school graduate |
|---|---|---|---|
| 2 | Nursery school to 4th grade | 10 | Some college, but less than 1 year |
| 3 | 5th grade or 6th grade | 11 | One or more years of college, no degree |
| 4 | 7th grade or 8th grade | 12 | Associate degree |
| 5 | 9th grade | 13 | Bachelor's degree |
| 6 | 10th grade | 14 | Master's degree |
| 7 | 11th grade | 15 | Professional degree |
| 8 | 12th grade, no diploma | 16 | Doctorate degree |

Dividing original group $N_1$ into 3 clusters we obtained 1365 families (16.45%) in the first cluster, 4538 families (54.68%) in the second cluster, and 2396 couples (28.87%) in the third one.

The next step of the Influence search algorithm is to define subsets of the group $N_2$ corresponding to clusters of group $N_1$. First it is necessary to choose parameters, which determine clusters and are invariant for both groups.

Among all parameters which were used for clustering only parents' age satisfies these requirements. All other parameters can change their values depending on state government policy (for example, young families can be provided with cheap education loans or material assistance).

In order to find age ranges which define clusters' bounds (according to the second part of the proposed algorithm fourth step) we calculated the percentage of representatives of a certain age in each cluster. Values corresponding to 80 or more percents of the appropriate number series maximum value were considered as cluster-defining bounds (Fig. 2). As a result we got age ranges 30-34 for men and 28-32 for women as the first cluster bounds, 27-30 and 24-31 as the second cluster bounds, 32-36 and 29-34 as the third cluster bounds.

These values were used to define cluster prototypes out of the group $N_2$. We got the first cluster prototype with 2862

elements, the second cluster prototype with 3430 elements, and the third one with 2789 elements.

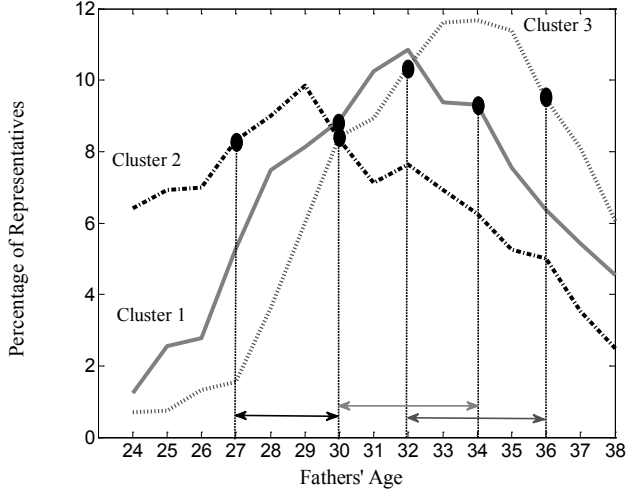

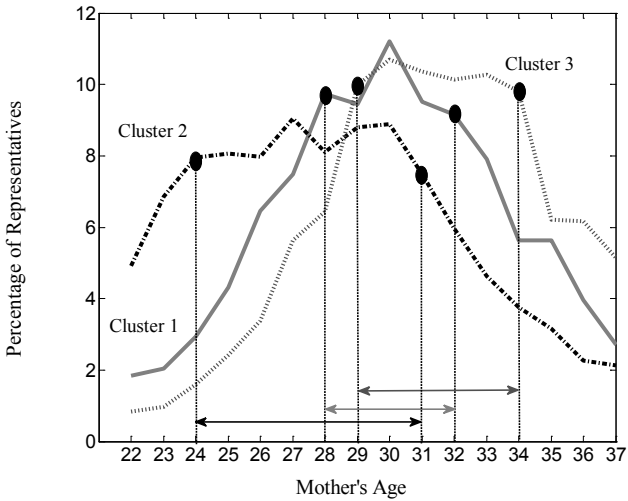

Fig. 2 Father's (a) and mother's (b) age distribution within clusters.

The next stage of the experiment was to compare characteristics of clusters from group $N_1$ and their prototypes from group $N_2$. After the appropriate calculations were held it turned out that significant differences take place only for the following attributes: father's total income, parents' education, home ownership, and type of building.

Fig. 3 shows fathers' total income distribution within each cluster and appropriate prototype. Corresponding income intervals are given in Table III. Additionally Fig. 3 shows that representatives from the second cluster have the lowest incomes, from the third cluster – the highest, and families from the first cluster correspond to intermediate values. By comparing income distributions in clusters and their prototypes we can say that a certain percentage of families from prototypes have lower incomes than families from the appropriate clusters. Especially significant difference occurs between the third cluster and its prototype.

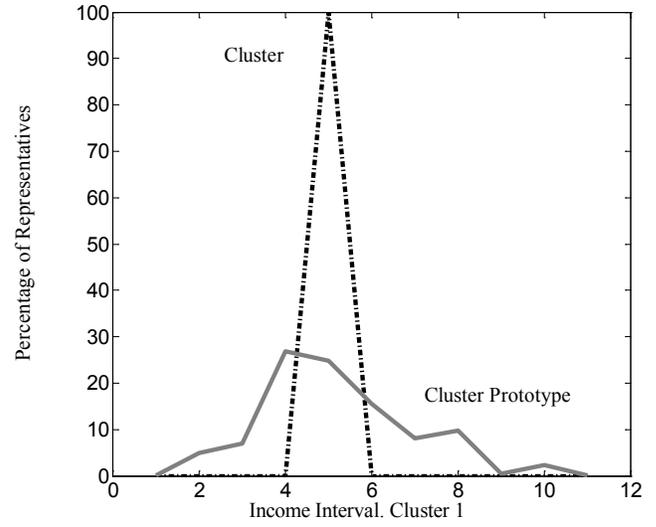

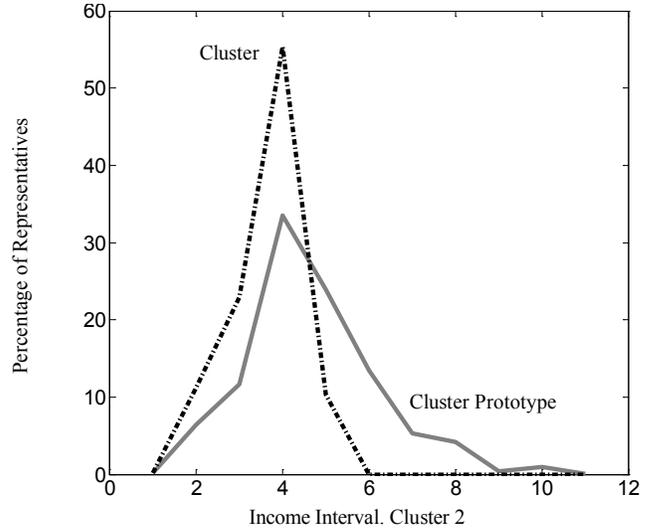

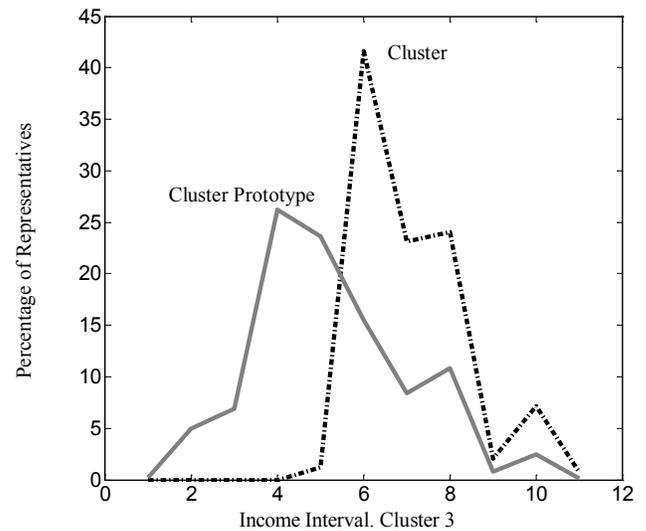

Fig. 3 Father's income distribution in the first (a), second (b), third (c) clusters and their prototypes.

TABLE III
FATHER'S TOTAL INCOME INTERVALS ( *10^4 )

| 1 | [-1; 0) | 7 | [8; 10) |
|---|---|---|---|
| 2 | [0; 1) | 8 | [10; 20) |
| 3 | [1; 2) | 9 | [20; 30) |
| 4 | [2; 4) | 10 | [30; 40) |
| 5 | [4; 6) | 11 | [40; 72] |
| 6 | [6; 8) | | |

Fig. 4 describes both father's and mother's education distribution. By analyzing it we can say that the second cluster contains families with the least educated parents, the third one – the most educated, and the first one occupies the intermediate position once again. At the same time, education levels in first and second clusters are lower than values in their prototypes. The third cluster has inverse dependence. It should be noted that the second cluster prototype, which consists of the youngest couples, is the source of replenishment for the first and the third clusters. Indeed, with the course of time families with highly educated parents can move to one of these clusters depending on when the first baby (babies) appears. Similarly, families from the first cluster prototype can move to the third cluster.

Home ownership distribution is shown on Fig. 5. The meaning of signs on a figure is following: 1 – owned by someone in this household with a mortgage or loan; 2 – owned by someone in this household free and clear (without a mortgage or loan); 3 – rented for cash rent; 4 – occupied without payment of cash rent.

It turns out that the vast majority of families either buys a dwelling on credit or rents it. We can note that in the first and the third clusters the greater part of families pays credits for their own homes and in their prototypes much of couples rent it. As for the second cluster the percentage of owners with a loan and leaseholders in cluster and its prototype are almost equal.

Concerning distribution among clusters we can say that as age increases (as well as income and education level) the percentage of owners with a loan grows and the percentage of leaseholders goes down.

Fig. 6 shows building type distribution. We used following signs: 1 – a mobile home; 2 – a one-family house detached from any other house; 3 – a one-family house attached to one or more houses; 4 – a building with 2..50 apartments.

The general distribution tendency as well as in cases considered above is rather natural: while moving from the second cluster to the first and then to the third one the percentage of families living in detached houses increases and the percentage of couples living in apartments goes down. Meanwhile figure shows that all prototypes have higher rates of families living in apartments and lower rates of couples occupying detached houses than appropriate clusters. It indicates lack of own dwelling.

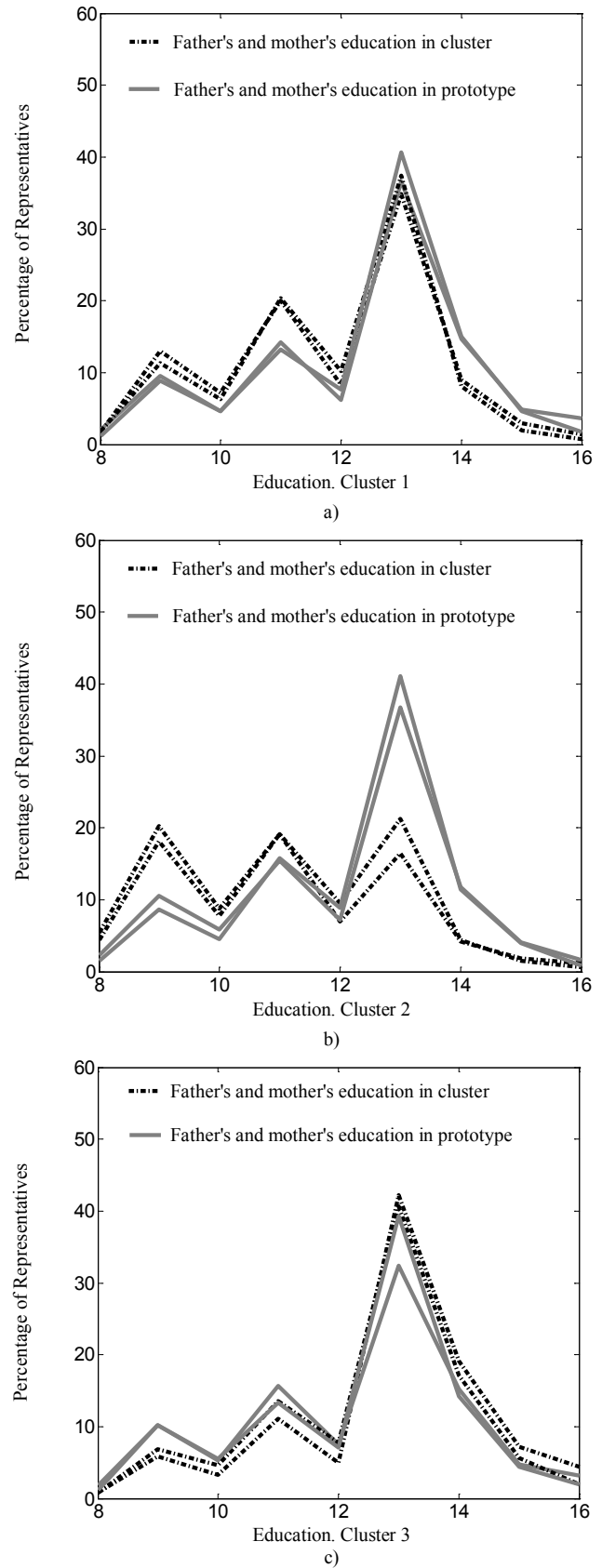

Fig. 4 Education distribution in the first (a), second (b), third (c) clusters and their prototypes.

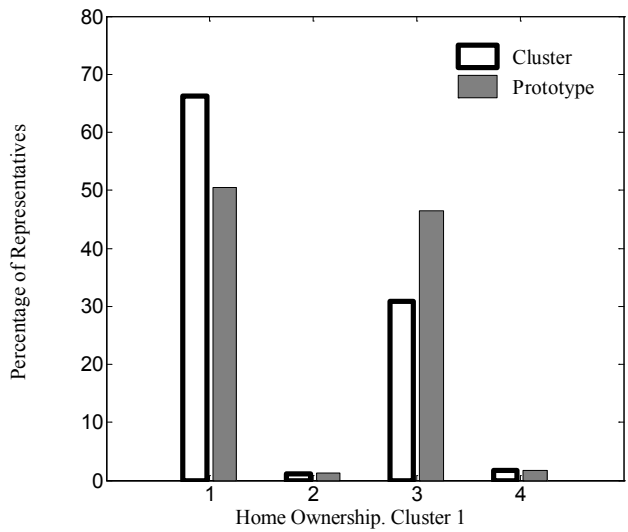

a)

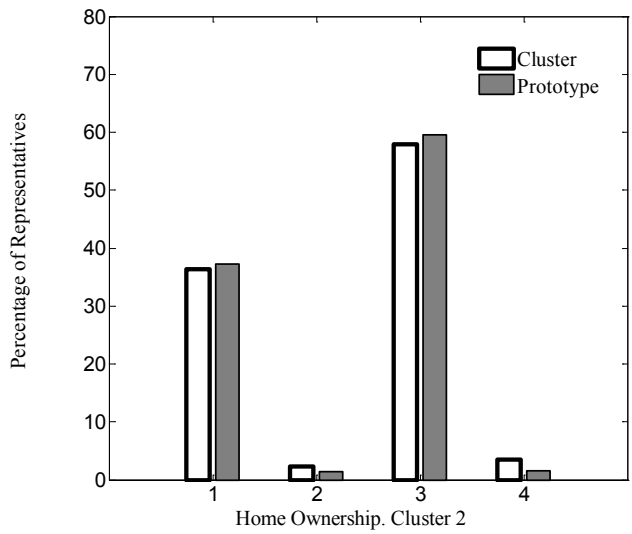

b)

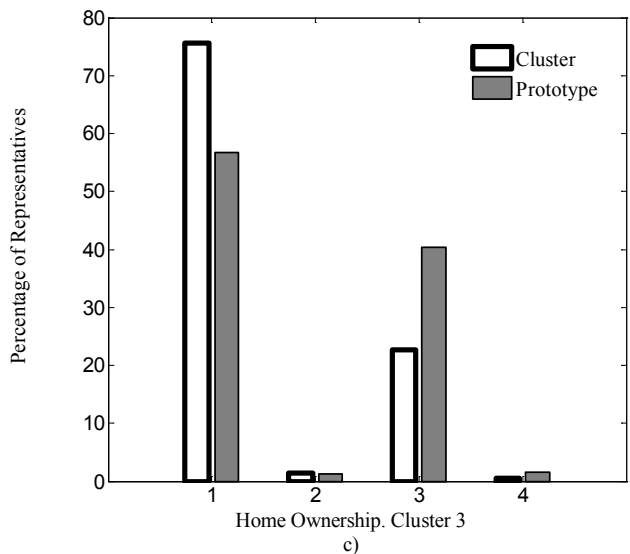

c)

Fig. 5 Home ownership distribution in the first (a), second (b), third (c) clusters and their prototypes.

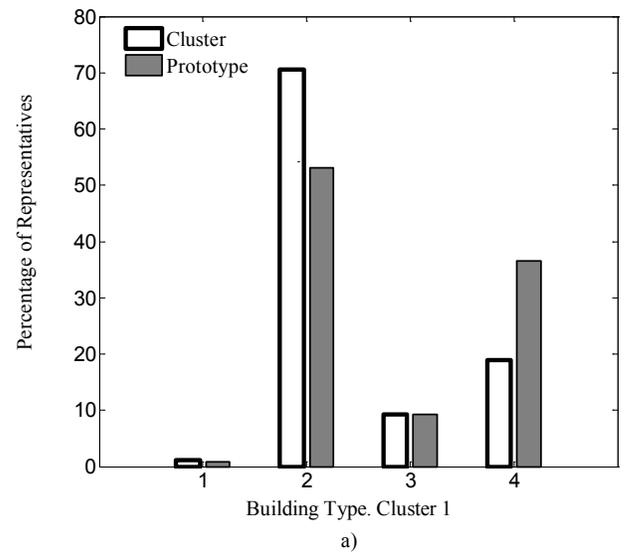

a)

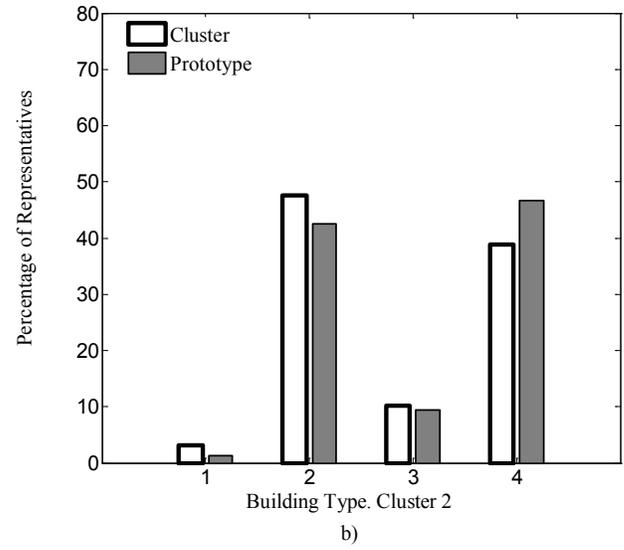

b)

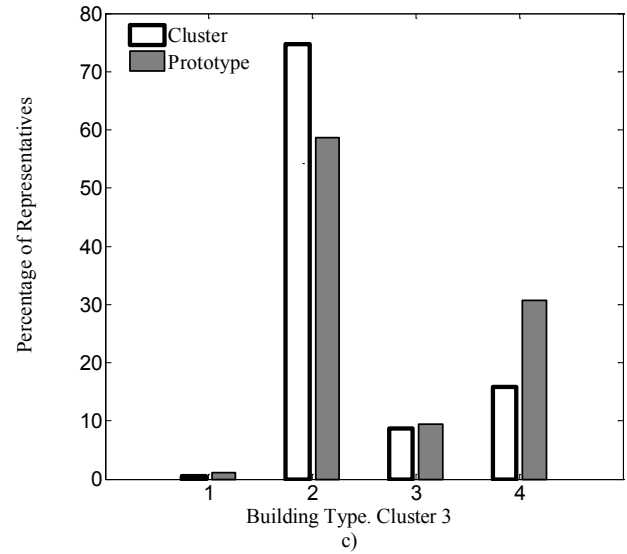

c)

Fig. 6 Building type distribution in the first (a), second (b), third (c) clusters and their prototypes.

Summing up experimental results we can say the following:
- providing financial support or cheep housing loans increases birth-rate;
- providing financial support to the youngest age group representatives with high education level probably won't contribute to their desire having babies, because most highly educated couples decide to have a baby after both parents are 30 years old;
- if we want couples from the oldest age group to have a baby, we should actively encourage them materially, a lot of families from this group lack money or own dwelling; besides, if the first child is not born before parents reach 40 years, the probability of his appearance reduces significantly.

## IV. Conclusions

Current paper shows that using Data Mining techniques for statistical data analysis, in particular for census data analysis, we can get results unreachable by other analysis methods. However, using only Data Mining may not be enough. Analysis algorithms which include Data Mining techniques as elements, just like the algorithm proposed in this paper, are of particular interest.

Developed Influence search algorithm is based on clustering a set which elements possess a certain feature, and defining clusters' prototypes out of the set with elements without this feature. Comparing characteristics of clusters and their prototypes we can give some recommendations regarding question how to "move" respondents from the second set to the first one (i.e., add this feature to respondents). Methodologically this algorithm resembles a well-known re-engineering approach, when we move from the "as-is" model to the model "to-be".

Among prospects for further researches there is a joint analysis of such population influencing factors as natality and migration. We also believe that some interesting results can be obtained out of a similar analysis for different US states and countries.